# Hyper Zagreb Index of Bridge and Chain Grpahs

Nilanjan De
Calcutta Institute of Engineering and Management, Kolkata, India. E-mail: de.nilanjan@rediffmail.com

**ABSTRACT**

*Let G be a simple connected molecular graph with vertex set V(G) and edge set E(G). One important modification of classical Zagreb index, called hyper Zagreb index HM(G) is defined as the sum of squares of the degree sum of the adjacent vertices, that is, sum of the terms $[d_G(u) + d_G(v)]^2$ over all the edges of G, where $d_G(u)$ denote the degree of the vetex u of G. In this paper, the hyper Zagreb index of certain bridge and chain graphs are computed and hence using the derived results we compute the hyper Zagreb index of several classes of chemical graphs and nanostructures.*

## 1. INTRODUCTION

In theoretical chemistry different molecular structures are often modelled using molecular graphs. Molecular graphs are actually a graphical representation of molecular structure through vertices and edges so that each vertex corresponds to atoms and the edges represents the bonds between them. Graph theory provides an importent tool called molecular graph-based structure-descriptor or more commonly topological index to correlate the physico-chemical properties of chemical compounds with their molecular structure. A topological index is a numeric amount derived from a molecular graph which characterize the topology of the molecular graph and is invariant under automorphism. Thus a topological index *Top(G)* of a graph *G* is a number such that, for each graph *H* isomorphic to *G*, *Top(G)=Top(H)*. The concept of topological index was first originated by Wiener, while he was chipping away at breakin point of paraffin. In this paper, we are concerned with molecular graphs, that is, the graphs which are simple, connected and having no directed or weighted edges. Let *G* be such a graph with vertex set $V(G)$ and edge set $E(G)$. Let the number of vertices and edges of *G* will be denoted by *n* and *m* respectively. Also let the edge connecting the vertices *u* and *v* is denoted by *uv*. The degree of a vertex *v*, is the number of first neighbors of *v* and is denoted by $d_G(v)$. Among different types of topological indices, the first and second Zagreb indices are most important topological indices in study of structure property correlation of molecules and have received a lot of attention in mathematical as well as chemical literature. The First and Second Zagreb Index was first introduced by Gutman and Trinajstić [1] and are defined as

$$M_1(G) = \sum_{v \in V(G)} d_G(u)^2 = \sum_{uv \in E(G)} \left[d_G(u) + d_G(v)\right] \text{ and } M_2(G) = \sum_{u,v \in V(G)} d_G(u) d_G(v).$$

These two Zagreb indices are among the oldest molecular structure descriptors and have been extensively studied both with respect to mathematical and chemical point of view (see [2-7]).

Different modification, generalization and extension of Zagreb indices have been introduced and studied by many researchers. One such modified version of Zagreb index called hyper Zagreb index, was introduced by G.H. Shirdel, H. Rezapour and A.M. Sayadi in [8], which is defined as

$$HM(G) = \sum_{uv \in E(G)} [d(u) + d(v)]^2.$$

There are various study of this index in recent literature (see [9-12] ).

We know that, a bridge or a chain graphs is the composite graphs formed from d number of different graphs $\{G_i\}_{i=1}^d$ by connecting them in proper sence. Thus different relationship for various topological indices can be established between the whole molecular graph with their components $\{G_i\}_{i=1}^d$. Also, it is interestingly found that many important molecular graphs and nanostructure are composed of some isomorphic subgraphs. So the expressions of different topological indices can be obtained by considering some special cases of bridge and chain graphs. In this paper, we consider two types of bridge graphs and a chain graph. There are some other studies in this direction for some topological indices also. Azari et al. in [13] derived some explicit expression of first and second Zagreb indices of bridge and chain graphs. In [14], the present author studied forgotten topological index or F-index of bridge and chain graphs with there respective applications. Mansour and Schork in [15] and [16] respectvely calculated PI index, Szeged index, Wiener, hyper-Wiener, detour and hyper-detour indices of different bridge and chain graphs where as they studied vertex PI index and Szeged index of bridge graphs in [18]. Li et al. in [17] calculated the vertex PI index and Szeged index of chain graphs.

In this study, first we derive different expressions for bridge and chain graphs which are composition of $d$ number graphs denoted by $G_i$, for $i = 1, 2, ..., d$ which are not necessarily isomorphic. Hence, we calculate the relations when all $G_i$ are isomorphic. Finally using these relations to find hyper Zagreb index of some special chemical graphs and nanostructure.

## 2. MAIN RESULTS

Bridge and chain graphs are composite graphs formed from $d$ different graphs $\{G_i\}_{i=1}^d$ by joining or connecting them in different ways. In this paper we consider two types of bridge graphs and one chain graph. In this section first we define these two types of bridge graphs namely

$$B_1 = B_1(G_1, G_2, ..., G_d; v_1, v_2, ..., v_d)$$
$$B_2 = B_2(G_1, G_2, ..., G_d; v_1, w_1, v_2, w_2, ..., v_d, w_d)$$

and hence compute their hyper Zagreb index. Then we define a chain graph denoted by

$$C = C(G_1, G_2, ..., G_d; v_1, w_1, v_2, w_2, ..., v_d, w_d)$$

and hence calculate the same of this.

### 2.1. Hyper Zagreb index of the bridge graph

The bridge graph of $\{G_i\}_{i=1}^d$ with respect to the vertices $\{v_i\}_{i=1}^d$ is denoted by

$$B_1 = B_1(G_1, G_2, ..., G_d; v_1, v_2, ..., v_d)$$

which is the graph obtained from $G_1, G_2, ..., G_d$ by connecting the vertices $v_i$ and $v_{i+1}$ by an edge for all $i = 1, 2, ..., d-1$. The bridge graph $B_1$ is shown in the Figure 1. From definition of $B_1$ we get following Lemma directly.

**Lemma 1.** The degree of the vertices of $B_1$ is given by

$$d_{B_1}(u) = \begin{cases} d_{G_i}(u), & \text{if } u \in V(G_i) - \{v_i\} \\ v_1 + 1, & \text{if } u = v_1 \\ v_i + 2, & \text{if } u = v_i, 2 \leq i \leq d-1 \\ v_d + 1, & \text{if } u = v_d \end{cases}$$

where $v_i = d_{G_i}(v_i)$, for $1 \leq i \leq d$.

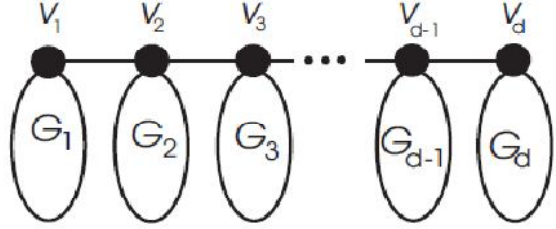

*Figure 1. The bridge graph $B_1 = B_1(G_1, G_2, ..., G_d; v_1, v_2, ..., v_d)$*

In the next theorem we now calculate the hyper Zagreb index of bridge graph of type $B_1$.

**Theorem 1.** The hyper Zagreb index of the bridge graph $B_1$, $d \geq 2$, is given by

$$HM(B_1) = \sum_{i=1}^{d} HM(G_i) + 6\sum_{i=2}^{d-1} v_i^2 + 20\sum_{i=3}^{d-2} v_i + 2\sum_{i=1}^{d-1} v_i v_{i+1} + 4\sum_{i=2}^{d-1} \delta_{G_i}(v_i) + 3(v_1^2 + v_d^2) + 7(v_1 + v_d)$$
$$+ 18(v_2 + v_{d-1}) + 2(\delta_{G_1}(v_1) + \delta_{G_d}(v_d)) + 16d - 30.$$

where $v_i = d_{G_i}(v_i)$, for $1 \leq i \leq d$.

**Proof.** From the construction of the bridge graph $B_1$, the edge set of $B_1$ is given by
$$E(B_1) = E(G_1) \cup E(G_2) \cup ... \cup E(G_d) \cup \{v_i v_{i+1} : 1 \leq i \leq (d-1)\}.$$

Now to calculate $HM(B_1)$, we use lemma 1 and partition the total sum into the following sums. First we consider the sum $S_1(B_1)$ over the edges $ab \in E(G_1)$.

$$S_1(B_1) = \sum_{ab \in E(G_1)} [d_{B_1}(a) + d_{B_1}(b)]^2$$
$$= \sum_{ab \in E(G_1); a,b \neq v_1} [d_{G_1}(a) + d_{G_1}(b)]^2 + \sum_{ab \in E(G_1); a \in V(G_1), b = v_1} [d_{G_1}(a) + (d_{G_1}(v_1) + 1)]^2$$
$$= \sum_{ab \in E(G_1); a,b \neq v_1} [d_{G_1}(a) + d_{G_1}(b)]^2 + \sum_{ab \in E(G_1); a \in V(G_1), b = v_1} [d_{G_1}(a) + d_{G_1}(v_1)]^2$$
$$+ \sum_{ab \in E(G_1); a \in V(G_1), b = v_1} [1 + 2(d_{G_1}(a) + d_{G_1}(v_1))]$$
$$= HM(G_1) + d_{G_1}(v_1) + 2\delta_{G_1}(v_1) + 2d_{G_1}(v_1)^2.$$

Next we consider the sum $S_2(B_1)$ over the edges $ab \in E(G_d)$.
$$S_2(B_1) = \sum_{ab \in E(G_d)} [d_{B_1}(a) + d_{B_1}(b)]^2$$

$$= \sum_{ab \in E(G_d); a,b \neq v_d} [d_{G_d}(a) + d_{G_d}(b)]^2 + \sum_{ab \in E(G_d); a \in V(G_d), b = v_d} [d_{G_d}(a) + (d_{G_d}(v_d) + 1)]^2$$

$$= \sum_{ab \in E(G_d); a,b \neq v_d} [d_{G_d}(a) + d_{G_d}(b)]^2 + \sum_{ab \in E(G_d); a \in V(G_d), b = v_d} [d_{G_d}(a) + d_{G_d}(v_d)]^2$$

$$+ \sum_{ab \in E(G_d); a \in V(G_d), b = v_d} [1 + 2(d_{G_d}(a) + d_{G_d}(v_d))]$$

$$= HM(G_d) + d_{G_d}(v_d) + 2\delta_{G_d}(v_d) + 2d_{G_d}(v_d)^2.$$

Similarly, the third sum $S_3(B_1)$ is taken over the edges $ab \in E(G_i)$, for $i = 2, 3, ..., (d-1)$.

$$S_3(B_1) = \sum_{i=2}^{d-1} \sum_{ab \in E(G_i)} [d_{B_1}(a) + d_{B_1}(b)]^2$$

$$= \sum_{i=2}^{d-1} \sum_{ab \in E(G_i); a,b \neq v_i} [d_{G_i}(a) + d_{G_i}(b)]^2 + \sum_{i=2}^{d-1} \sum_{ab \in E(G_i); a \in V(G_i), b = v_i} [d_{G_i}(a) + (d_{G_i}(b) + 2)]^2$$

$$= \sum_{i=2}^{d-1} \sum_{ab \in E(G_i); a,b \neq v_i} [d_{G_i}(a) + d_{G_i}(b)]^2 + \sum_{i=2}^{d-1} \sum_{ab \in E(G_i); a \in V(G_i), b = v_i} [d_{G_i}(a) + d_{G_i}(v_i)]^2$$

$$+ \sum_{i=2}^{d-1} \sum_{ab \in E(G_i); a \in V(G_i), b = v_i} [4 + 4(d_{G_i}(a) + d_{G_i}(v_i))]$$

$$= \sum_{i=2}^{d-1} HM(G_i) + 4 \sum_{i=2}^{d-1} d_{G_i}(v_i)^2 + 4 \sum_{i=2}^{d-1} d_{G_i}(v_i) + 4 \sum_{i=2}^{d-1} \delta_{G_i}(v_i).$$

Finally the contribution $S_4(B_1)$ of all the edges $v_i v_{i+1}$, for $i = 1, 2, ..., (d-1)$ is gfiven by.

$$S_4(B_1) = \sum_{i=1}^{d-1} \sum_{ab = v_i v_{i+1}} [d_{B_1}(a) + d_{B_1}(b)]^2$$

$$= [(d_{G_1}(v_1) + 1) + (d_{G_2}(v_2) + 2)]^2 + \sum_{i=2}^{d-2} [(d_{G_i}(v_i) + 2) + (d_{G_{i+1}}(v_{i+1}) + 2)]^2$$

$$+ [(d_{G_{d-1}}(v_{d-1}) + 2) + (d_{G_d}(v_d) + 1)]^2$$

$$= (v_1 + v_2 + 3)^2 + \sum_{i=2}^{d-2} (v_i + v_{i+1} + 4)^2 + (v_{d-1} + v_d + 3)^2$$

$$= 2\sum_{i=2}^{d-1} v_i^2 + 16 \sum_{i=3}^{d-1} v_i + 2 \sum_{i=1}^{d-1} v_i v_{i+1} + (v_1^2 + v_d^2) + 6(v_1 + v_d) + 14(v_2 + v_{d-1}) + 16d - 30.$$

Hence combining the contributions $S_1(B_1)$, $S_2(B_1)$, $S_3(B_1)$ and $S_4(B_1)$, we get the desired expression given in theorem 1. □

Let, $v$ is a vertex of a graph $G$, and $G_i = G$ and $v_i = v$ for $i = 1, 2, ..., d$, then using above theorem we directly get the following :

**Corollary 1.** The hyper Zagreb index of the bridge graph $B_1 = B_1(G, G, ., G; v, v, ., v)$, $d \geq 2$, is given by

$$HM(B_1) = dHM(G) + 8(d-1)v^2 + 10(2d-3)v + 4(d-1)\delta_G(v) + 16d - 30$$

where $v = d_G(v)$.

Another kind of bridge graph of $\{G_i\}_{i=1}^d$ with respect to the vertices $\{v_i, w_i\}_{i=1}^d$ is denoted by

$$B_2 = B_2(G_1, G_2, ., G_d; v_1, w_1, v_2, w_2, ., v_d, w_d),$$

which is the graph obtained from $G_1, G_2, ..., G_d$ by connecting the vertices $w_i$ and $v_{i+1}$ by an edge for all $i = 1, 2, ..., d-1$. The bridge graph $B_2$ is shown in the Figure 2. Similarly the following Lemma is the direct consequence of the definition of $B_2$.

**Lemma 2.** The degree of the vertices of $B_2$ are given by

$$d_{B_2}(u) = \begin{cases} d_{G_1}(u), & \text{if } u \in V(G_1) - \{w_1\} \\ d_{G_d}(u), & \text{if } u \in V(G_d) - \{v_d\} \\ d_{G_i}(u), & \text{if } u \in V(G_i) - \{v_i, w_i\}, 2 \le i \le d-1 \\ w_i + 1, & \text{if } u = w_i, 1 \le i \le d-1 \\ v_i + 1, & \text{if } u = v_i, 2 \le i \le d \end{cases}$$

where $v_i = d_{G_i}(v_i)$, $w_i = d_{G_i}(w_i)$, for $1 \le i \le d$.

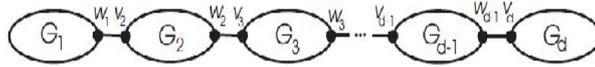

*Figure 2. The bridge graph $B_2 = B_2(G_1, G_2, ..., G_d; v_1, w_1, v_2, w_2, ..., v_d, w_d)$*

In the following theorem we now calculate hyper Zagreb index of the bridge graph $B_2$.

**Theorem 2.** The hyper Zagreb index of the bridge graph $B_2$, $d \ge 2$, is given by

$$HM(B_2) = \sum_{i=1}^{d} HM(G_i) + \sum_{i=1}^{d-1}[3w_i^2 + 5w_i + 2\delta_{G_i}(w_i)] + \sum_{i=1}^{d-1}[3v_i^2 + 5v_i + 2\delta_{G_i}(v_i)] + 2\sum_{i=1}^{d-1}v_i w_i + 4(d-1)$$

where $v_i = d_{G_i}(v_i)$, $w_i = d_{G_i}(w_i)$; $v_i$ and $w_i$ are not adjacent, for $1 \le i \le d$.

**Proof.** From the construction of the bridge graph $B_2$, the edge set of $B_2$ is given by
$$E(B_2) = E(G_1) \cup E(G_2) \cup ... \cup E(G_d) \cup \{v_i v_{i+1} : 1 \le i \le (d-1)\}.$$
Now to calculate $HM(B_2)$, we use lemma 1 and partition the total sum into the following sums. First we consider the sum $S_1(B_2)$ over the edges $ab \in E(G_1)$.

$$S_1(B_2) = \sum_{ab \in E(G_1)} [d_{B_2}(a) + d_{B_2}(b)]^2$$

$$= \sum_{ab \in E(G_1); a,b \ne w_1} [d_{G_1}(a) + d_{G_1}(b)]^2 + \sum_{ab \in E(G_1); a \in V(G_1), b=w_1} [d_{G_1}(a) + (d_{G_1}(w_1)+1)]^2$$

$$= \sum_{ab \in E(G_1); a,b \ne w_1} [d_{G_1}(a) + d_{G_1}(b)]^2 + \sum_{ab \in E(G_1); a \in V(G_1), b=w_1} [d_{G_1}(a) + d_{G_1}(w_1)]^2$$

$$+ \sum_{ab \in E(G_1); a \in V(G_1), b=w_1} [1 + 2(d_{G_1}(a) + d_{G_1}(w_1))]$$

$$= HM(G_1) + d_{G_1}(w_1) + 2\delta_{G_1}(w_1) + 2d_{G_1}(w_1)^2.$$

Similarly, we have the sum $S_2(B_2)$ over the edges $ab \in E(G_d)$.

$$S_2(B_2) = \sum_{ab \in E(G_d)} [d_{B_2}(a) + d_{B_2}(b)]^2$$

$$= \sum_{ab\in E(G_d);a,b\neq v_d} [d_{G_d}(a)+d_{G_d}(b)]^2 + \sum_{ab\in E(G_d);a\in V(G_d),b=v_d} [d_{G_d}(a)+(d_{G_d}(v_d)+1)]^2$$

$$= \sum_{ab\in E(G_d);a,b\neq v_d} [d_{G_d}(a)+d_{G_d}(b)]^2 + \sum_{ab\in E(G_d);a\in V(G_d),b=v_d} [d_{G_d}(a)+d_{G_d}(v_d)]^2$$

$$+ \sum_{ab\in E(G_d);a\in V(G_d),b=v_d} [1+2(d_{G_d}(a)+d_{G_d}(v_d))]$$

$$= HM(G_d)+d_{G_d}(v_d)+2\delta_{G_d}(v_d)+2d_{G_d}(v_d)^2.$$

Similarly, the third sum $S_3(B_2)$ is taken over the edges $ab \in E(G_i)$, for $i = 2,3,...,(d-1)$.

$$S_3(B_2) = \sum_{i=2}^{d-1} \sum_{ab\in E(G_i)} [d_{B_2}(a)+d_{B_2}(b)]^2$$

$$= \sum_{i=2}^{d-1} \sum_{ab\in E(G_i);a,b\neq v_i,w_i} [d_{G_i}(a)+d_{G_i}(b)]^2 + \sum_{i=2}^{d-1} \sum_{ab\in E(G_i);a\neq w_i} [d_{G_i}(a)+(d_{G_i}(b)+1)]^2$$

$$+ \sum_{i=2}^{d-1} \sum_{ab\in E(G_i);a\neq v_i} [d_{G_i}(a)+(d_{G_i}(b)+1)]^2 + \sum_{i=2}^{d-1} [(d_{G_i}(v_i)+1)+(d_{G_i}(w_i)+1)]^2$$

$$= \sum_{i=2}^{d-1} \sum_{ab\in E(G_i);a,b\neq v_i,w_i} [d_{G_i}(a)+d_{G_i}(b)]^2 + \sum_{i=2}^{d-1} \sum_{ab\in E(G_i);a\neq w_i} [d_{G_i}(a)+d_{G_i}(v_i)]^2$$

$$+ \sum_{i=2}^{d-1} \sum_{ab\in E(G_i);a\neq v_i} [d_{G_i}(a)+d_{G_i}(w_i)]^2 + \sum_{i=2}^{d-1} \sum_{ab\in E(G_i);a\neq w_i} [1+2(d_{G_i}(a)+d_{G_i}(v_i))]$$

$$+ \sum_{i=2}^{d-1} \sum_{ab\in E(G_i);a\neq v_i} [1+2(d_{G_i}(a)+d_{G_i}(w_i))]$$

$$= \sum_{i=2}^{d-1} HM(G_i) + \sum_{i=2}^{d-1} [2d_{G_i}(v_i)^2+d_{G_i}(v_i)+2\delta_{G_i}(v_i)+2d_{G_i}(w_i)^2+d_{G_i}(w_i)+2\delta_{G_i}(w_i)].$$

Finally, the contribution $S_4(B_2)$ of all the edges $w_i v_{i+1}$, for $i = 2,3,...,(d-1)$ is calculated as follows.

$$S_4(B_2) = \sum_{i=2}^{d-1} \sum_{ab=w_i v_{i+1}} [d_{B_1}(a)+d_{B_1}(b)]^2$$

$$= \sum_{i=2}^{d-2} [(d_{G_i}(w_i)+1)+(d_{G_i}(v_{i+1})+1)]^2$$

$$= \sum_{i=2}^{d-2} (w_i+v_{i+1}+4)^2$$

$$= \sum_{i=2}^{d-1} [w_i^2+v_{i+1}^2+2w_i v_{i+1}+4w_i+4v_{i+1}+4].$$

Now combining all the contributions $S_1(B_2)$, $S_2(B_2)$, $S_3(B_2)$ and $S_4(B_2)$, we get the desired expression given in theorem 2. □

Let, $u$ and $v$ are two vertices of a graph $G$, and $G_i = G$, $v_i = v$ and $w_i = w$ for all $i = 1,2,...,d$. Then using above theorem we directly get the following.

**Corollary 2.** The hyper Zagreb index of the bridge graph $B_2$, $d \geq 2$, is given by

$$F(B_2) = dF(G)+3(d-1)(v^2+w^2+v+w)+2(d-1)$$

where $v = d_G(v)$, $w = d_G(w)$ and $v$, $w$ are not adjacent.

## 2.2. Hyper Zagreb index of Chain Graph

The chain graph $\{G_i\}_{i=1}^d$ with respect to the vertices $\{v_i, w_i\}_{i=1}^d$ is denoted by
$$C = C(G_1, G_2, ..., G_d; v_1, w_1, v_2, w_2, ..., v_d, w_d),$$
which is the graph obtained from $G_1, G_2, ..., G_d$ by identifying the vertices $w_i$ and $v_{i+1}$ for all $i = 1, 2, ..., d-1$. The chain graph is shown in the Figure 3.

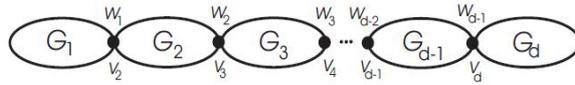

Figure 3. The chain graph $C = C(G_1, G_2, ..., G_d; v_1, w_1, v_2, w_2, ..., v_d, w_d)$

From definition of chain graph C, we can state the following Lemma.

**Lemma 3.** The degree of the vertices of chain graph $C$ are given by
$$d_C(v) = \begin{cases} d_{G_1}(v), & \text{if } v \in V(G_1) - \{w_1\} \\ d_{G_d}(v), & \text{if } v \in V(G_d) - \{v_d\} \\ d_{G_i}(v), & \text{if } v \in V(G_i) - \{v_i, w_i\}, 2 \leq i \leq d-1 \\ w_i + v_{i+1}, & \text{if } u = w_i = v_{i+1}, 1 \leq i \leq d-1 \end{cases}$$
where $v_i = d_{G_i}(v_i)$, $w_i = d_{G_i}(w_i)$, for $1 \leq i \leq d$.

In the following theorem we now calculate hyper Zagreb index of chain graph C.

**Theorem 3.** The hyper Zagreb index of the Chain graph C, $d \geq 2$, is given by
$$HM(C) = \sum_{i=1}^{d} HM(G_i) + \sum_{i=2}^{d-1}[w_{i-1}^2 v_i + 2w_{i-1}\delta_{G_i}(v_i) + 2w_{i-1}v_i^2] + \sum_{i=1}^{d-1}[w_i v_{i+1}^2 + 2v_{i+1}\delta_{G_i}(w_i) + 2v_{i+1}w_i^2]$$
$$+ w_1 v_2^2 + 2v_2 \delta_{G_1}(w_1) + 2v_2 w_1^2 + w_{d-1}^2 v_d + 2w_{d-1}\delta_{G_d}(v_d) + 2w_{d-1}v_d^2.$$
where $v_i = d_{G_i}(v_i)$, $w_i = d_{G_i}(w_i)$; $v_i$ and $w_i$ are not adjacent, for $1 \leq i \leq d$.

**Proof.** Using definition of F-index and Lemma 3, we get
First, we have the sum $S_1(C)$ over the edges $ab \in E(G_1)$ as follows.
$$S_1(C) = \sum_{ab \in E(G_1)} [d_C(a) + d_C(b)]^2$$
$$= \sum_{ab \in E(G_1); a,b \neq w_1} [d_{G_1}(a) + d_{G_1}(b)]^2 + \sum_{ab \in E(G_1); a \in V(G_1), b = w_1} [d_{G_1}(a) + (d_{G_1}(w_1) + d_{G_2}(v_2))]^2$$
$$= \sum_{ab \in E(G_1); a,b \neq w_1} [d_{G_1}(a) + d_{G_1}(b)]^2 + \sum_{ab \in E(G_1); a \in V(G_1), b = w_1} [d_{G_1}(a) + d_{G_1}(w_1)]^2$$
$$+ \sum_{ab \in E(G_1); a \in V(G_1), b = w_1} [d_{G_2}(v_2)^2 + 2d_{G_2}(v_2)(d_{G_1}(a) + d_{G_1}(w_1))]$$
$$= HM(G_1) + d_{G_2}(v_2)^2 d_{G_1}(w_1) + 2d_{G_2}(v_2)\delta_{G_1}(w_1) + 2d_{G_2}(v_2)d_{G_1}(w_1)^2.$$

Similarly, we have the sum $S_2(C)$ over the edges $ab \in E(G_d)$.

$$S_2(C) = \sum_{ab \in E(G_d)} [d_C(a) + d_C(b)]^2$$

$$= \sum_{ab \in E(G_d); a,b \neq v_d} [d_{G_d}(a) + d_{G_d}(b)]^2 + \sum_{ab \in E(G_d); a \in V(G_d), b = v_d} [d_{G_d}(a) + (d_{G_{d-1}}(w_{d-1}) + d_{G_d}(v_d))]^2$$

$$= \sum_{ab \in E(G_d); a,b \neq v_d} [d_{G_d}(a) + d_{G_d}(b)]^2 + \sum_{ab \in E(G_d); a \in V(G_d), b = v_d} [d_{G_d}(a) + d_{G_d}(v_d)]^2$$

$$+ \sum_{ab \in E(G_d); a \in V(G_d), b = v_d} [d_{G_{d-1}}(w_{d-1})^2 + 2d_{G_{d-1}}(w_{d-1})(d_{G_d}(a) + d_{G_d}(v_d))]$$

$$= HM(G_d) + d_{G_{d-1}}(w_{d-1})^2 d_{G_d}(v_d) + 2d_{G_{d-1}}(w_{d-1})\delta_{G_d}(v_d) + 2d_{G_{d-1}}(w_{d-1})d_{G_d}(v_d)^2.$$

The third sum $S_3(C)$ is taken over the edges $ab \in E(G_i)$, for $i = 1, 2, ..., (d-1)$.

$$S_3(C) = \sum_{i=2}^{d-1} \sum_{ab \in E(G_i)} [d_C(a) + d_C(b)]^2$$

$$= \sum_{i=2}^{d-1} \sum_{ab \in E(G_i); a,b \neq v_i, w_i} [d_{G_i}(a) + d_{G_i}(b)]^2 + \sum_{i=2}^{d-1} \sum_{ab \in E(G_i); a \neq w_i, b = v_i} [d_{G_i}(a) + (d_{G_{i-1}}(w_{i-1}) + d_{G_i}(v_i))]^2$$

$$+ \sum_{i=2}^{d-1} \sum_{ab \in E(G_i); a \neq v_i, b = w_i} [d_{G_i}(a) + (d_{G_i}(w_i) + d_{G_{i+1}}(v_{i+1}))]^2$$

$$+ \sum_{i=2}^{d-1} [(d_{G_{i-1}}(w_{i-1}) + d_{G_i}(v_i)) + (d_{G_i}(w_i) + d_{G_{i+1}}(v_{i+1}))]^2$$

$$= \sum_{i=2}^{d-1} \sum_{ab \in E(G_i); a,b \neq v_i, w_i} [d_{G_i}(a) + d_{G_i}(b)]^2 + \sum_{i=2}^{d-1} \sum_{ab \in E(G_i); a \neq w_i} [d_{G_i}(a) + d_{G_i}(v_i)]^2$$

$$+ \sum_{i=2}^{d-1} \sum_{ab \in E(G_i); a \neq v_i} [d_{G_i}(a) + d_{G_i}(w_i)]^2$$

$$+ \sum_{i=2}^{d-1} \sum_{ab \in E(G_i); a \neq w_i} [d_{G_{i-1}}(w_{i-1})^2 + 2d_{G_{i-1}}(w_{i-1})(d_{G_i}(a) + d_{G_i}(v_i))]$$

$$+ \sum_{i=2}^{d-1} \sum_{ab \in E(G_i); a \neq v_i} [d_{G_{i+1}}(v_{i+1})^2 + 2d_{G_{i+1}}(v_{i+1})(d_{G_i}(a) + d_{G_i}(w_i))]$$

$$= \sum_{i=2}^{d-1} HM(G_i) + \sum_{i=2}^{d-1} [2d_{G_i}(v_i)^2 + d_{G_i}(v_i) + 2\delta_{G_i}(v_i) + 2d_{G_i}(w_i)^2 + d_{G_i}(w_i) + 2\delta_{G_i}(w_i)].$$

Now combining the sums $S_1(C)$, $S_2(C)$ and $S_3(C)$ the desired result follows. □

Let, $u$ and $v$ are two vertices of a graph $G$, and $G_i = G$, $v_i = v$ and $w_i = w$ for $i = 1, 2, ..., d$. Then using above theorem we directly get the following.

**Corollary 3.** The hyper Zagreb index of the chain graph $C$, $d \geq 2$, is given by
$$HM(C) = dHM(G) + (d-1)[3vw(v+w) + 2(w\delta_G(v) + v\delta_G(w))]$$
where $v = d_G(v)$, $w = d_G(w)$ and $v, w$ are not adjacent.

## 3. APPLICATION : HYPER ZAGREB INDEX OF SOME GRAPHS

As many chemically important molecular graphs and nano structures are composed of some isomorphic components, so in this section we study such graphs which consists of some isomorphic components and can be treated as a particular case of above discussed bridge and chain graphs. Thus, here we apply the results derived in the previous section to compute hyper Zagreb index of some special chemically interesting molecular graphs and also of nanostructures. Let us first consider application of first type of bridge graph $B_1$.

**Example 1**. First we consider a particular bridge graph, denoted by
$$T_{d,n} = G_d(C_n, v) = B_1(C_n, C_n, ..., C_n; v, v, ..., v) \text{ (}d\text{-times)},$$
where $v$ is an arbitrary vertex of degree 2 of $C_n$ (see Figure 4). Applying Corollary 1, after direct calculation we get the following
$$HM(T_{d,n}) = 16nd + 104d - 138.$$

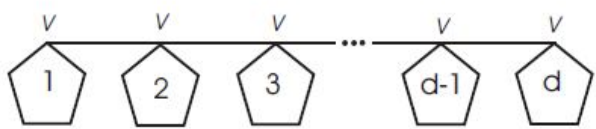

*Figure 4. The bridge graph $B_1(C_5, C_5, ..., C_5; v, v, ..., v)$*

**Example 2**. Let us now consider another particular bridge graph, denoted by
$$B_d = G_d(P_3, v) = B_1(P_3, P_3, ..., P_3; v, v, ..., v) \text{ (}d\text{-times)},$$
where $v$ is the middle vertex of degree 2 of the path graph $P_3$ (see Figure 5). Applying Corollary 1, after direct calculation we get the following
$$HM(B_d) = 114d - 130.$$

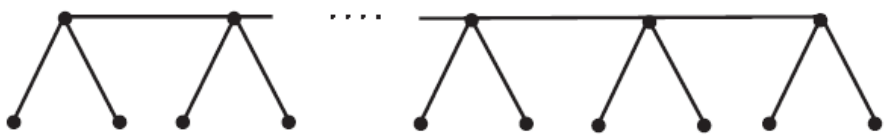

*Figure 5. The bridge graph $B_1(P_3, P_3, ..., P_3; v, v, ..., v)$*

**Example 3.** Let us now consider a particular bridge graph, denoted by
$$A_{d,m} = G_d(P_m, v) = B_1(P_m, P_m, ..., P_m; v, v, ..., v) \text{ (d-times)},$$
where $v$ is the vertex of degree 1 of $P_m$ (see Figure 6). Applying Corollary 1, after direct calculation we get the following
$$HM(A_{d,m}) = 16md + 22d - 76.$$

Note that, the square comb lattice graph $C_q(N)$ with open ends, is the bridge graph $A_{n,n}$, where $N = n^2$ is the total number of vertices.

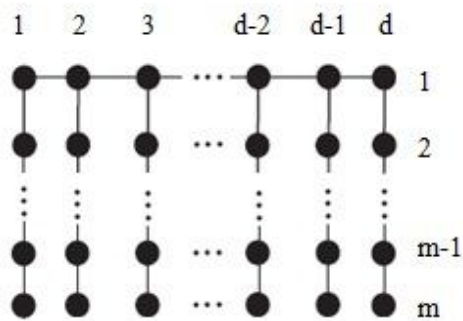

*Figure 6. The bridge graph graph $A_{d,m}$*

**Example 4.** Now we consider a particular bridge graph called the van Hove comb lattice graph $CvH(N)$ with open ends. This graph can be represented as the bridge graph $B_1(P_1, P_2, ..., P_{n-1}, P_n, P_{n-1}, ..., P_2, P_1; v_{1,1}, v_{1,2}, ..., v_{1,n-1}, v_{1,n}, v_{1,n-1}, ..., v_{1,2}, v_{1,1})$, where for $2 \leq i \leq n$, $v_{1,i}$ is the first vertex of degree one of the path graph $P_i$ and $v_{1,1}$ is the single vertex of degree zero of the path $P_1$ (Figure 7). So, using Corollary 1 we get, after calculation

$$HM(CvH(N)) = 16n^2 + 44n - 106.$$

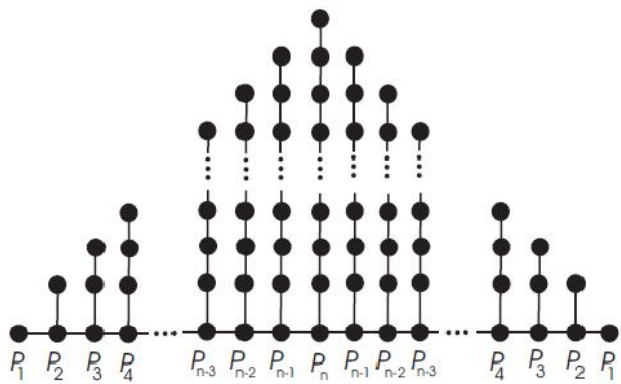

*Figure 7. The van Hove comb lattice graph $CvH(N)$*

**Example 5.** Next consider the molecular graph of the nanostar dendrimers $D_n$ (as shown in Figure 8). This graph is equivalent to the bridge graph $B_2(G, G, ..., G; v, w, v, w, ..., v, w)$ (n-times), where $G$ is the graph given in Figure 9 and $v$ and $w$ are vertices of degree 2. Thus applying corollary 2, we get

$$HM(D_n) = 450n + (n-1)72 = 522n - 72.$$

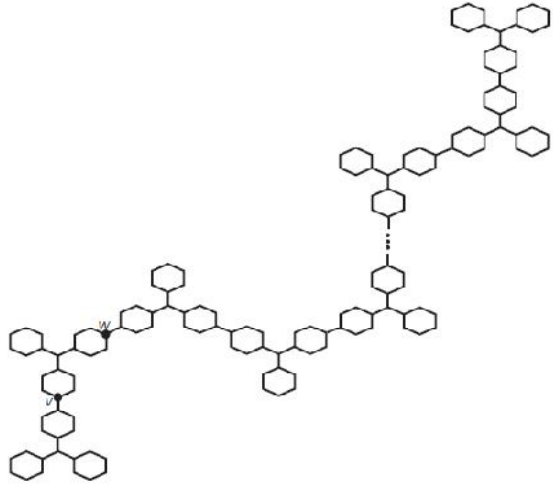

*Figure 8. The molecular graph of the nanostar dendrimers $D_n$.*

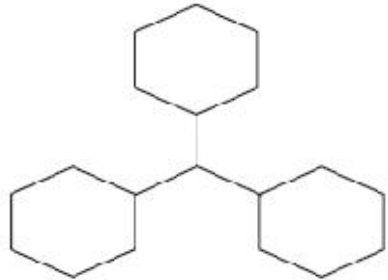

*Figure 9. The molecular graph of the nanostar dendrimers $D_1$*

**Example 6.** The polyphenyl chain of $h$ hexagon is said to be an ortho-($O_h$), meta-($M_h$) and para-($L_h$), if all its internal hexagons are ortho-hexagons, meta-hexagons and para-hexagons respectively. The polyphenyl chain $O_h$, $M_h$ and $L_h$ may be considered as the second type of bridge graph

$$B_2(C_6, C_6, ..., C_6; v, w, v, w, ..., v, w).$$

Here $C_6$ is the cycle graph where every vertex is of degree two (i.e. $v = w = 2$). The ortho-, meta-, para- polyphenyl chain of hexagons is given in Figure 10. So, using Corollary 2, we get from direct calculation

$$HM(M_h) = HM(L_h) = 96h + (h-1)72 = 168h - 72.$$

Also, applying Corollary 1, we get

$$HM(O_h) = 200h - 138.$$

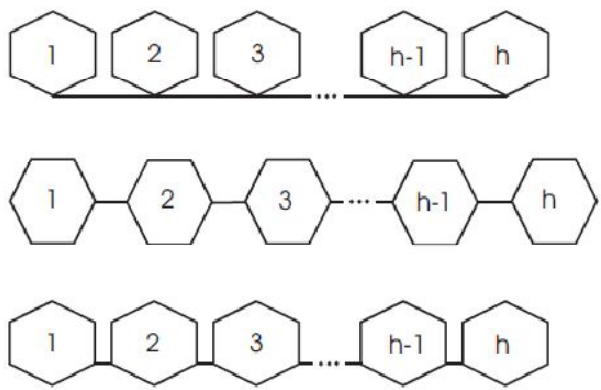

*Figure 10. The ortho-, meta-, para- polyphenyl chain of hexagons*

**Example 7.** The spiro-chain of the graph $G = C_n(k,l)$ $(n \geq 3)$ is the chain graph $C(G,G,...,G; v,w,v,w,...,v,w)$ where $k$ and $l$ are numbers of vertices $v$ and $w$ respectively. The spiro-chain of $C_4$ and $C_6$ are shown in Figure 11. The spiro-chain of $d$ number of $C_n(k,l)$ is denoted by $S_d(C_n(k,l))$. Thus applying Corollary 3, we get

$$HM(S_d(C_n(k,l))) = 16nd + 80d - 80.$$

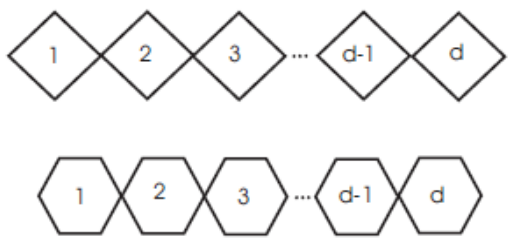

*Figure 11. The spiro- chain of $C_4$ and $C_6$*

## 4. CONCLUSION

In this paper, we have derived explicit expressions of hyper Zagreb index of different bridge and chain graphs. Also, as a special case of bridge and chain graphs, we have calculated expressions of hyper Zagreb index of several chemically important graphs and nanostructures, which are built from several copies of isomorphic graphs.